\documentclass[letter]{aa} % for the letters
\usepackage[utf8]{inputenc}
\usepackage[T1]{fontenc}

\usepackage{ulem}
\usepackage{enumitem}
\usepackage{amsmath}
\usepackage{comment}

\usepackage{graphicx}
\usepackage{multirow} 
\usepackage{amssymb}
\usepackage{color}
\usepackage[switch]{lineno}
\usepackage{url}
\graphicspath{{../figure/}}
\usepackage{setspace}
\newcommand\U[1]{{\,\rm #1}}
\newcommand\al{\alpha}

\newcommand\gm{\gamma}

\newcommand\Dl{\Delta}
\newcommand\eps{\epsilon}

\newcommand\rs[1]{_\mathrm{#1}}

\newcommand\Edot{\dot E}

\newcommand\Omc{\Omega\rs{c}}

\newcommand\jp{j\rs{p}}
\newcommand\kmax{k\rs{max}}
\newcommand\vA{v\rs{A}}
\newcommand\tsync{\tau\rs{sync}}

\newcommand\Vpsr{V\rs{psr}}
\newcommand\gmMPD{\gm\rs{\,MPD}}

\newcommand\gmX{\gm\rs{\,X}}
\newcommand\gmesc{\gm\rs{esc}}
\newcommand\epslim{\eps\rs{lim}}

\newcommand\repo[1]{{\color{black}#1}}

\usepackage{natbib,twoopt}
\usepackage[breaklinks=true]{hyperref} %% to avoid \citeads line fills
\bibpunct{(}{)}{;}{a}{}{,}             %% natbib format for A&A and ApJ
\makeatletter
  \newcommandtwoopt{\citeads}[3][][]{\href{http://adsabs.harvard.edu/abs/#3}%
    {\def\hyper@linkstart##1##2{}%
     \let\hyper@linkend\@empty\citealp[#1][#2]{#3}}}
  \newcommandtwoopt{\citepads}[3][][]{\href{http://adsabs.harvard.edu/abs/#3}%
    {\def\hyper@linkstart##1##2{}%
     \let\hyper@linkend\@empty\citep[#1][#2]{#3}}}
  \newcommandtwoopt{\citetads}[3][][]{\href{http://adsabs.harvard.edu/abs/#3}%
    {\def\hyper@linkstart##1##2{}%
     \let\hyper@linkend\@empty\citet[#1][#2]{#3}}}
  \newcommandtwoopt{\citeyearads}[3][][]%
    {\href{http://adsabs.harvard.edu/abs/#3}
    {\def\hyper@linkstart##1##2{}%
     \let\hyper@linkend\@empty\citeyear[#1][#2]{#3}}}
\makeatother
\title{The nature of the X-ray filaments around bow shock pulsar wind nebulae}
\titlerunning{The nature of the X-ray filaments around BSPWNe}
\authorrunning{B.Olmi et al.}

\author{Barbara Olmi\inst{1,2}, Elena Amato\inst{1,3}, Rino Bandiera\inst{1} and Pasquale Blasi\inst{4,5}}
\institute{INAF - Osservatorio Astrofisico di Arcetri, Largo E. Fermi 5, I-50125 Firenze, Italy
\and
INAF - Osservatorio Astronomico di Palermo, Piazza del Parlamento 1, I-90134 Palermo, Italy
\and 
Università degli Studi di Firenze, Via Sansone 1, 50019, Sesto Fiorentino (FI), Italy
\and
GSSI - Gran Sasso Science Institute, Viale F. Crispi 7 - I-67100 L’ Aquila, Italy
\and
INFN-Laboratori Nazionali del Gran Sasso, Via G. Acitelli 22, Assergi (AQ), Italy\\
\email{barbara.olmi@inaf.it,elena.amato@inaf.it,pasquale.blasi@gssi.it}
}

\abstract
{We propose that the X-ray filaments emerging from selected bow shock pulsar wind nebulae are due to a charge-separated outflow of electrons and/or positrons escaping the nebula and propagating along the local Galactic magnetic field.}
{The X-ray brightness, length, and thickness of filaments are all accounted for if a nonresonant streaming instability is excited. }
{This is possible if particles are released in the interstellar medium as a collimated beam, as would be expected in a reconnection region between the nebular and interstellar magnetic fields. }
{We successfully test this idea on the Guitar Nebula filament and discuss other cases.
}
{These filaments provide the best diagnostics available for particle escape from evolved pulsar wind nebulae, a process essential to assessing the contribution of these sources to cosmic ray positrons. The same phenomenology might govern the occurrence of TeV halos and their importance for cosmic ray transport.}

\let\linenumbers\nolinenumbers\nolinenumbers
\begin{document}
%
%\flushbottom
\maketitle
\thispagestyle{empty}
\section{Introduction}
An increasing number of bow shock pulsar wind nebulae (BSPWNe) shows evidence of filamentary X-ray structures \citep{Hui_Becker:2007,Pavan:2014,Temim:2015,Klingler:2016,Klingler:2018,Medvedev:2019,Marelli:2019,Bordas_Zhang:2020,Zhang:2020,Klingler_Yang:2020,deVries-J2030:2022} protruding in the interstellar medium (ISM).
The best characterized examples are the Guitar Nebula \citep{Hui_Becker:2007,deVries:2022}, PSR J2030+4415 \citep{deVries:2020,deVries:2022}, and the Lighthouse Nebula \citep{Pavan:2014,Pavan:2016,Klingler:2023}. 
The filaments appear elongated in one direction for a distance ranging from $0.6\U{pc}$ in the case of the Guitar Nebula to $\sim 15\U{pc}$ in the case of the Lighthouse, with relatively mild morphological fluctuations. 
The filaments' thickness is very small \repo{(from less than $1\%$ of the length to $\sim 10\%$)}. 
The spectrum and morphology of the X-ray emission are compatible with synchrotron radiation of very high-energy leptons in a magnetic field that is substantially larger than the typical interstellar magnetic field (typically by a factor of $\sim 10$), a possible indication \citep{Bandiera:2008} that some type of instability is excited by particles escaping the nebula \citep{Olmi:2023}, leading to magnetic field amplification. In turn, this can account for the thickness of the filaments as a consequence of severe synchrotron energy losses. 
This qualitatively appealing picture needs a quantitative connection with known instabilities and with the actual energetics of the particles leaving the BSPWN. 

Here we show that the length and  thickness of the filaments require specific conditions that can only be satisfied if 1) electrons and positrons with sufficiently high energies are spatially charge-separated when leaving the BSPWN \citep{Olmi_Bucciantini:2019c,Olmi_Bucciantini:2023}, and 2) they are numerous enough to excite the nonresonant hybrid instability  \citep{Bell:2004} (NRI). 
Neither of these conditions is trivial: charge separation can only occur for pairs with energy sufficiently close to the maximum potential drop (MPD) of the pulsar, a condition that severely constrains the current density available to excite the instability.

We make the case that the electrons (or positrons) that escape the BSPWN may be focused in a narrow angle when leaving the source, and propagate along the local Galactic magnetic field (GMF) lines. The angular collimation is important in that it confines particles in a region with a small cross section, which in turn increases the current density. 
The excitation requires that the energy density associated with the particles dominating the current be larger than that of the preexisting magnetic field \citep{Bell:2004} (in this case the GMF).

The nonresonant nature of the instability is a crucial ingredient in this context: the lack of scattering allows the current-carrying particles to stream away from the BSPWN at speed close to that of light ($c$), thereby filling a filament for a length of order $\sim c\tau_{CR}$, where $\tau_{CR}$ is the time needed for the saturation of the instability. During this phase, the perturbed magnetic field keeps growing, possibly beyond the average Galactic value, on scales much smaller than the Larmor radius of the particles dominating the current. At saturation, or close to it, the nonlinear evolution of the instability drives power on large scales  \citep{Bell:2004}, and eventually on scales comparable to the Larmor radius of the particles in the amplified field. At this point, particles start scattering efficiently and rapidly isotropize. The synchrotron emission of these electrons or positrons in the amplified field is expected to produce the X-rays that we observe from a region of length $\sim c\tau_{CR}$.

The proper motion of the pulsar is characterized by the pulsar velocity, $\Vpsr$, and the 
(typically large) angle, $\theta_f$, that it forms with the local GMF (which is assumed to coincide in direction with the filament elongation). This would shape the emission region as a wide stripe, unless the emitting particles suffer synchrotron losses on a timescale, $\tau\rs{cross}=w_f/(\Vpsr\sin\theta_f)$, with $w_f$ being the filament width.

%The proper motion of the pulsar in a direction that is in general different from - and typically at a large angle ($\theta_f$) to - that of the filament, would shape the emission region as a wide stripe, unless the emitting particles suffer synchrotron losses on a time-scale $\tau\rs{cross}=w_f/(\Vpsr\sin\theta_f)$, with $w_f$ the filament width and $\Vpsr$ the pulsar velocity.
% 
Imposing $\tau\rs{loss}=\tau\rs{cross}$ returns the strength of the magnetic field in the filament, %\LEt{***It is a bit ambiguous what is being compared here. Please clarify.}
that must be compared with the value inferred from the saturation of the NRI. We show that the picture we are proposing is able to describe in a coherent way the length, thickness, and luminosity of the observed filaments. 

\begin{comment}
The interpretation of the filaments in terms of excitation of the NRI from charge separated electrons/positrons released by BSPWNe into the ISM is rich of implications. The amplified turbulence level can lead to extended confinement of the particles in selected regions around the sources, with implications on the pulsar contribution to CR leptons. It is tantalizing to speculate that the recently detected TeV-halos, regions of Very High Energy ($\gtrsim \mathrm{TeV}$) gamma-ray emission extending for tens of pc around a few pulsars \citep{TeVHalos2017}, might be due to a similar phenomenology, in that they also require particle transport to be suppressed by $2-3$ orders of magnitude \citep{TeVHaloRecv} compared with the Galactic values as inferred from CR secondary/primary ratios  \citep{Schroer2023,Evoli2020}.
\end{comment}

%%%%%%%%%%%%%%%%%%%%%%%%%%%%%%%%%%%%%%%%%%%%%%%%%%
\section{\repo{Physical model of the filament}}\label{sec:results}
%
%\BOC{[1419 parole]}\\

The discussion that follows aims at a quantitative description of three observables: (i) the length of the feature, $L_f$, (ii) its transverse thickness, $w_f\ll L_f$, and
(iii) the observed X-ray luminosity in a given energy band, $L_{\mathrm{X}_1-\mathrm{X}_2}$. 
To date, the filaments have only been reliably measured in a handful of sources: the Guitar Nebula, the Lighthouse, and J2030+4415. These will be our benchmark cases.
As was discussed early on \citep{Bandiera:2008}, the small filament widths require that the magnetic field inside these structures be larger than the GMF (typically by a factor of $\sim10$) so that synchrotron losses are fast enough. 
At the same time, however, the radiating particles must reach distances from the pulsar much larger than the thickness of the filaments. 
This latter requirement suggests that the instability responsible for magnetic amplification is nonresonant. A resonant instability would in fact limit the particle motion to a few Larmor radii from the pulsar, due to effective resonant scattering, which is incompatible with the observed length. Here we show that the excitation of the current-driven NRI \citep{Bell:2004} satisfies all these conditions and provides a suitable explanation of the observed phenomena. 

While BSPWNe are expected to produce an equal number of electrons and positrons, it has been shown that particles with different charges escape the nebula along different paths \citep{Olmi_Bucciantini:2019c} at energies $\lesssim m c^2 \gmMPD$, where \repo{the Lorentz factor corresponding to the MPD is}
\begin{equation}\label{eq:gmpd}
    \gmMPD = \frac{e}{m c^2} \sqrt{\frac{\dot{E}}{c}}\,.
\end{equation}
$m$ and $e$ are the electron mass and charge, respectively, and $\dot{E}$ is the spin down luminosity of the pulsar. 
The current associated with the escaping particles can be expressed as $j\rs{p}=e\,\eps \dot{E}/(m\,c^2\,\gmesc \, A)$, where $\eps$ is the fraction of $\dot{E}$ carried by the escaping particles, $\gmesc$ is the minimum energy of the escaping particle, and $A=\pi R^2$ is the area at the base of the filament.
%
%This expression for $n_p$ assumes a spectrum  for escaping particles: $N(\gamma)\propto \gamma^{-p}$ with $p>2$, so that most of the energy is carried by particles around $\gmesc$. This condition is fulfilled since typically $\gmesc>\gamma_b$, where $\gamma_b\sim 5 \times 10^5$ is the Lorentz factor where the spectral index $p$ changes from $p=1.1-1.6$ to $p=2.2-2.5$ \citep{Bucciantini:2011,Torres:2014}. 
%
The current can then be written as
\begin{equation}\label{eq:j}
    \jp  = \eps \,\frac{m\,c^3}{e\pi R^2}\,\frac{\gmMPD^2}{\gmesc}.
\end{equation}
The NRI excites perturbations that in the linear and early nonlinear phases grow on small scales, with the maximum growth occurring at
\begin{equation}
    \label{eq:kmax}
    \kmax=\frac{4\pi}{B_0c}\jp=
    \frac{4 \eps}{R^2} \, \frac{c}{\Omc} \, \frac{\gmMPD^2}{\gmesc}, 
\end{equation}
at a rate,
\begin{equation}
    \label{eq:GmBell}
    \tau_{CR}^{-1}\quad=\kmax\vA=  \frac{4 \eps}{R^2} \, \frac{c^2}{\Omc}\, \frac{\vA}{c} \frac{\gmMPD^2}{\gmesc},
\end{equation}
where $\vA=B_0/\sqrt{4\pi\,\rho_\mathrm{ISM}}$ is the Alfvén speed in the ISM with mass density \repo{$\rho_\mathrm{ISM}$} and $\Omc=eB_0/(mc)$ is the electron cyclotron frequency in the unperturbed ambient field, $B_0$.

The condition for these modes to be excited is $\kmax \; R_L>1$, with $R_L=m\,c^2\gmesc/(e\,B_0)$ the Larmor radius of the particles with $\gmesc$.
The inequality can be rewritten as 
\begin{equation}\label{eq:epslim}
    \eps > \epslim = \frac{R^2 \Omc^2}{4 c^2 \,\gmMPD^2} = \frac{1}{4}\,\frac{R^2}{R_{L,\,\mathrm{MPD}}^2}\,,
\end{equation}
where $R_{L,\,\mathrm{MPD}}=mc^2\gmMPD/(eB_0)$.
Excitement of the NRI requires the energy density carried by the electric current to locally exceed that
%\LEt{***or perhaps you mean “electric current locally to exceed that”} 
of the ambient magnetic field. 
This is reflected in a lower limit on the fraction of $\dot{E}$
%pulsar spin down luminosity 
that needs to be channelled into the tube: once $\eps>\epslim$, the NRI is excited.
The saturation of the NRI has been the subject of much literature: as was already found in the seminal work on this instability \citep{Bell:2004}, the growth of the unstable modes, associated with scales much smaller than $R_L$, is also accompanied by power deposition at larger scales, until there is power on scales comparable to the Larmor radius of the particles dominating the current calculated in the amplified field, $\Dl B$. 
At that point, scattering becomes important and the current is disrupted. 
This condition reads as \repo{$\kmax^*\,R_L^*=1$}, where $\kmax^*= 4\pi\jp/(c\Dl B)$ and $R_L^*=m\,c^2\gmesc/(e\,\Dl B)$, and it leads to
\begin{equation}\label{eq:bsat}
    \left(\frac{\Dl B}{B_0}\right)^2 = \frac{\eps}{\epslim} \,.
\end{equation}
%Several other physical interpretations of what happens at saturation have been put forward \RB{(REFS ???)}, resulting in a saturation field similar to that in Eq. \ref{eq:bsat}, that we will use in the following.\marginpar{\elena{se non si aggiungono referenze non serve}} 

It is important to keep in mind that during the exponential growth of the instability the perturbations remain on scales much smaller than the resonant scale, so that scattering is inhibited and, in the first approximation, the motion of the particles can be pictured as quasi-ballistic.
%, \RB{with a speed $\sim c \cos{\al}$, where $\alpha$ is the pitch angle of the particles producing the instability.}\RBcomm{Questo (che c'era già) è in dissonanza con quanto detto prima} 
We see a posteriori that a rather small collimation angle, $\alpha$, of the injected particles is required and the longitudinal speed of the particles in the quasi-ballistic phase is $\sim c$. This phase lasts for a few e-folds of the instability, $\sim 5$  \citep{Bell:2004}
, so that particles move along the magnetic field for a length,
\begin{eqnarray}\label{eq:featLength}
    L_f & \approx & 5c\,\tau_{CR} \;=\;
    \frac{5c}{\Omc}
    \left(\frac{c}{v_A}\right)
    \left(\frac{\eps}{\epslim}\right)^{-1}
    \gmesc \\ \nonumber
    &\simeq &
    422\U{pc} \left(\frac{\eps}{\epslim}\right)^{-1} 
    \gm_{\;7} \; n_{1}^{1/2} \; B_3^{-2}
    \,,
\end{eqnarray}
where 
%$\tau_{CR}$ is the growth time of NRI and 
in the last equality we have introduced  $\gm_{\;7}=\gmesc/10^7$,  $B_3=B_0/3\U{\mu G}$, and $n_{1}=\rho_\mathrm{ISM}/(1\, m_p\U{cm^{-3}})$, with $m_p$ the proton mass.

Within the assumed scenario, one can use Eq.~\ref{eq:bsat} and Eq.~\ref{eq:featLength} to relate the strength of the amplified magnetic field to the measured length of the filament. The result is
\begin{equation}\label{eq:closing1}
\Dl B 
%\simeq 52\U{\mu G}   \;\; n_{0.5}^{1/4} \;\gm_{\; 7}^{1/2} \left(\frac{L_f}{\mathrm pc}\right)^{-1/2}
\simeq 62\U{\mu G}   \;\; n_{1}^{1/4} \;\gm_{\; 7}^{1/2} \left(\frac{L_f}{\mathrm{ pc}}\right)^{-1/2}\,,
\end{equation}
which, as expected for the NRI, is independent of the initial value of the field, $B_0$.

A crucial ingredient in building a physical interpretation of the filaments is the motion of the pulsar with velocity $\Vpsr$. In the synchrotron loss time the pulsar has travelled a distance,
\begin{eqnarray}\label{eq:closing2}
     d\rs{sync}(\gamma) &=& \Vpsr \, \sin\theta_f \, \tau\rs{sync}(\gamma) =  \\ \nonumber
     &=& 5.7 \times 10^{-2} \; \mathrm{pc} \;\; \sin\theta_f \;  n_{1}^{-3/8}\; \gm_{\; 7}^{-3/4} \\ \nonumber
     && E\rs{X,keV}^{-1/2} \,\left( \frac{\Vpsr}{500\;\mathrm{km\,s}^{-1}}\right) \,  \left(\frac{L_f}{\mathrm{pc}}\right)^{3/4}. \nonumber
 \end{eqnarray}
In our scenario, $d\rs{sync}(\gamma_X)$ is interpreted as the measured thickness of the filament, $w_f$, and 
it allows one to derive $\gmesc$:
\begin{eqnarray}\label{eq:gesc}
    \gmesc &=& 
    4.5\times 10^5  \;\sin\theta_f^{4/3}\;   n_{1}^{-1/2}\; E\rs{X,keV}^{-2/3}  \\ \nonumber
    && \left( \frac{\Vpsr}{500\,\mathrm{km\,s}^{-1}}\right)^{4/3} \,\left(\frac{L_f}{\mathrm{pc}}\right) \; \left(\frac{w_f}{\mathrm{ pc}}\right)^{-4/3}, \nonumber
\end{eqnarray}
which turns out to depend only on directly observed quantities, except for the density. Lacking better constraints, the latter can be derived by measuring the standoff distance, $d_0$, defined by the balance between the pulsar wind energy flux and the ISM ram pressure: $\dot E/(4 \pi c d_0^2)=\rho_\mathrm{ISM}\Vpsr^2$.
%
%\begin{equation}\label{eq:density}
%n_0=6.7 \times 10^{-6}\U{cm^{-3}} \left(\frac{\dot{E}}{10^{35} \,\mathrm{erg \,s}^{-1} }\right) \left(\frac{d_0}{\rm pc}\right)^{-2}
%\left(\frac{\Vpsr}{500\U{km \,s}^{-1}}}\right)^{-2}\ .
%\end{equation}
We can then %\BO{use Eq.\ref{eq:density} in Eq.\ref{eq:closing2} and} 
express the Lorentz factor of the current-carrying particles as a function of observed quantities alone. This closes the system of equations.

\begin{table*}%[ht]
\centering
\fontsize{9}{9}\selectfont
\begin{tabular}{|l|c|c|c|c|c|c|c|c|c|}
\hline
System & $\dot{E}$ & $\tau_c$ & $d$  & $\Vpsr$ & $d_0$ & $L_{\mathrm{X}_1-\mathrm{X}_2}$  & $L_f$ & $w_f$ & $\sin\theta_f$\\
 & erg/s    &     kyr  &   kpc &  km/s  &   $^{\arcsec}$   &    erg/s (keV)  &             pc ($^{\;\prime}\,$) &   pc ($\,^{\arcsec}\,$) &  \\
\hline
Guitar$^{1,\,2,\,3}$ & $1.27\times 10^{33}$ & $1130$ &0.83 & $765.1$ &  0.06 & $3.6\times 10^{30}$ (0.5-7)  & 0.6 (2.5) & 0.08 (20) & 115 \\
\hline
Lighthouse$^{4,\,5,\,6}$ & $1.36\times 10^{36}$ & $116$ & 7 & $800-2400$ &  0.34 & $8.9\times 10^{33}$ (0.5-8)  & 15 (7.5) & 0.5 (16) & 118 \\
 &  &  & & & & $1.9\times 10^{34}$ (3-79)   &  &  & \\
\hline
J2030+4415$^7$ & $2.2\times10^{34}$ & $600$ & 0.72$^*$ &  $290.6$ & $0.3$ & $7.5\times 10^{30}$ (0.5-7)  & 3.14 (15) & 0.013$^{\dagger}$ (3.7) & 130  \\
\hline
\end{tabular}

\caption{\label{tab:infosystems}Relevant observational data of our benchmark systems. The main references are indicated next to the source name. $\dot E$ is the pulsar's spin-down power; $\tau_c$ its characteristic age; $d$ its distance; and $\Vpsr$ its projected velocity. $d_0$ is the bow shock standoff distance; $L_{\mathrm{X}_1-\mathrm{X}_2}$ indicates the X-ray luminosity of the feature in the energy range specified in parentheses; and $L_f$, $w_f$, and $\sin\theta_f$ are the feature's length, width, and inclination, respectively.
For the Lighthouse Nebula, we also report the X-ray luminosity in the 3-79 keV band recently measured by NuSTAR \citep{Klingler:2023}.\\
\footnotesize{$^*$ The distance is taken from the ATNF catalogue: \url{www.atnf.csiro.au/people/pulsar/psrcat}.\\
\footnotesize{$^{\dagger}$ The filament thickness appears to broaden with distance from the pulsar \citep{deVries-J2030:2022}. Limiting the estimate to the first third of the filament, one finds that $w_f=2^{\arcsec}$, a thickness comparable with Chandra's resolution.}\\
\footnotesize{$^1$\citet{deVries:2022}, $^2$\citet{Deller:2019}, $^3$\citet{Chatterjee_Cordes:2002}
}\\
\footnotesize{$^4$\citet{Pavan:2014}, $^5$\citet{Pavan:2016}, $^6$\citet{Klingler:2023}}\\
\footnotesize{$^7$\citet{deVries-J2030:2022}.}
}
%%% LEGEND MAX 350 PAROLE!
}
\end{table*}

Finally, we can make use of the observed X-ray luminosity to prove that the global picture presented above is meaningful. 
If the spectrum of the escaping particles reflects the power law behaviour that is typically observed in PWNe \citep{Torres:2014}, one expects that if 
$\gmesc\geq\gamma_b\sim 5\times 10^5$,  
then the number of injected particles per unit energy and time interval can be written as
$Q(\gamma)=Q_0 (\gamma/\gamma_b)^{-p}$, where $p\simeq 2.2-2.5$ is the high-energy particle injection index and $Q_0$ is a normalization constant that can be obtained by assuming that a fraction, $\eta$, of the power emitted by the pulsar goes into particles with $\gamma\geq\gamma_b$:
\begin{eqnarray}\label{eq:kappa}
    \nonumber
    &&\eta \Edot =m^2 c^4\int^{\gmMPD}_{\gamma_b} Q_0 \left(\frac{\gamma}{\gamma_b}\right)^{-p} \gamma \, d\gamma  \\
    && \rightarrow \quad 
    Q_0=  \frac{ p-2}{\left[1-(\gamma_b/\gmMPD)^{\,p-2}\right]} \frac{\eta \dot E}{(m\ c^2\ \gamma_b)^2}\,. 
\end{eqnarray}
Similarly, the energy injected in the form of particles with $\gamma \geq \gmesc$, namely the particles that escape the nebula and end up in the feature, is
\begin{eqnarray}\label{eq:injrate}
    \nonumber
    &&\eps\dot{E}=\int^{\gmMPD}_{\gmesc} Q_0\left(\frac{\gamma}{\gamma_b}\right)^{-p} \gamma \, d\gamma \\ 
    &&\rightarrow \quad 
    \eps = \eta \left(\frac{\gamma_b}{\gmesc}\right)^{p-2}
    \left(
    \frac{1-\left(\gmesc/\gmMPD\right)^{p-2}}
    {1-\left(\gamma_b/\gmMPD\right)^{p-2}}\right)\, ,
\end{eqnarray}
where we used Eq.~\ref{eq:kappa} for $Q_0$.
Under our assumptions, these particles will radiate all their energy in the feature, so that the X-ray luminosity in the energy range $E_{\mathrm{X}_1}-E_{\mathrm{X}_2}$ can be written as
\begin{eqnarray}\label{eq:Lx}
     L_{\mathrm{X}_1-\mathrm{X}_2}&=&(m\ c^2)^2 \int^{\gamma_{\mathrm{X}_2}}_{\gamma_{\mathrm{X}_1}} Q(\gamma) \gamma d\gamma = \\ \nonumber 
     &=&\eps \dot E 
     \left[\frac{1-\left(\gamma_{\,\mathrm{X}_1}/\gamma_{\,\mathrm{X}_2}\right)^{p-2}}{1-\left(\gmesc/\gmMPD\right)^{p-2}}\right] 
     \left(\frac{\gmesc}{\gamma_{\mathrm{X}_1}}\right)^{p-2}\,,
 \end{eqnarray}
where $\gm_{\mathrm{X}_i}\geq \gmesc$ is the Lorentz factor of a lepton that, in the amplified magnetic field, $\Delta B$, emits synchrotron radiation in the observed energy interval.
The assumption underlying Eq.\ref{eq:Lx} is that the life of the emitting particles is limited by synchrotron losses.
From the previous equation we then express our last unknown parameter, $\eps$, in terms of measured quantities alone:

\begin{equation}\label{eq:epsilon_general}
    \eps= \frac{L_{\mathrm{X}_1-\mathrm{X}_2}}{\dot{E}}
     \left[\frac{1-(\gmesc/\gmMPD)^{p-2}}{1-(\gamma_{\,\mathrm{X}_1}/\gamma_{\,\mathrm{X}_2})^{p-2}}\right] \left(\frac{\gamma_{\,\mathrm{X}_1}}{\gmesc}\right)^{p-2}\ . \end{equation}
This estimate assumes 
that the velocity distribution of particles is isotropic. As is shown in App.~\ref{sec:app1}, this is a reasonable approximation.

\repo{Once saturation of the NRI has been reached (after a time of $\sim 5\tau_{CR}$), the motion of the particles with Lorentz factor $\gmesc$ is no longer ballistic, since scattering causes the particles to diffuse, with a diffusion coefficient that can be estimated as
\begin{eqnarray}\label{eq:diff}
    D(\gmesc) &=&
    \frac{1}{3}\frac{m c^3 \gmesc}{e \Dl B} \\
    &\approx& 2.8\times 10^{24}\U{cm^2 \,s^{-1}} \; n_{1}^{-1/4} \; \gm_{\; 7}^{1/2} \left(\frac{L_f}{\mathrm{ pc}}\right)^{1/2}.\nonumber
\end{eqnarray}
For $\gm>\gmesc$, namely for more energetic particles that we assume to be responsible for the X-ray emission, the diffusion occurs in the small-scale turbulence regime, in which $D(\gm)=D(\gmesc)(\gm/\gmesc)^{2}$ \citep{Subedi:2017}.
%$D(\gm)\propto \gm^2$ or, more quantitatively \citep{Subedi:2017}, $D(\gm)=D(\gmesc)(\gm/\gmesc)^{2}$.
One can easily check that for the X-ray emitting particles the timescale for diffusive escape from the filament of length $L_f$ is much longer than the timescale for synchrotron losses:
\begin{eqnarray}
    \tau\rs{sync}(\gamma_X) &=& \frac{6\pi m c}{\sigma_T \Dl B^2 \gamma_X}\\
    &\approx& 
    115~\mathrm{yr}  \;\;  \; n_{1}^{-3/8} \; \gm_{\; 7}^{-3/4} \,  E\rs{X,keV}^{-1/2} \left(\frac{L_f}{\mathrm{pc}}\right)^{3/4},\nonumber
\end{eqnarray}
where $\sigma_T$ is the Thomson cross section and $\gamma_X$ the Lorentz factor of the X-ray emitting particle. In the second equality, we used the relation between $\gamma$ and peak synchrotron photon energy in the amplified field, and expressed the latter in keV as $E\rs{X,keV}=1.5\times10^{-17}B_3 \gamma_X^{\,2}$.}

%---------------------------------------------------
\section{Testing the scenario}
%\BOC{[461 parole]}\\
%
We could now test the proposed scenario on some filaments for which all of the quantities listed above were measured with sufficient accuracy. 
We used the measured parameters of the system, listed in Table~\ref{tab:infosystems}, to derive $\gmesc$ from Eq.~\ref{eq:gesc}. We could then use $\gmesc$ in Eq.~\ref{eq:closing1} and estimate the value of $\Delta B$. This also set the ratio, $\eps/\epslim$, through Eq.~\ref{eq:bsat} and, once $\eps$ was estimated from X-ray measurements through Eq.\ref{eq:epsilon_general}, we could finally constrain the size of the particle injection region, $R$, via Eq.~\ref{eq:epslim}. The results of this procedure are reported in Table~\ref{tab:results}.

The Guitar Nebula and its filament represent a prototypical case: this was the first system discovered \citep{Hui_Becker:2007}, and to date it is the best characterized, thanks to multi-epoch monitoring \citep{deVries:2022}. The Guitar Nebula is produced by the radio pulsar PSR B2224+65. 
Its bow shock is only visible in H$_\alpha$ emission \citep{Chatterjee_Cordes:2002}, and it has a peculiar guitar-like shape: in the $\sim 80^{\arcsec}$-long tail 
one can distinguish a thin, elongated head, followed by a wider body made by at least two bubbles. This peculiar shape has been interpreted as the result of mass loading of neutral atoms into the bow shock from a dense ambient medium \citep{Morlino:2015,Olmi:2018}.
As was expected, given its age, the pulsar has a rather low %spin-down power
$\dot{E}$, corresponding to a maximum Lorentz factor, $\gmMPD\simeq 1.2\times 10^8$.
\begin{figure*}
\centering
\includegraphics[width=.5\textwidth]{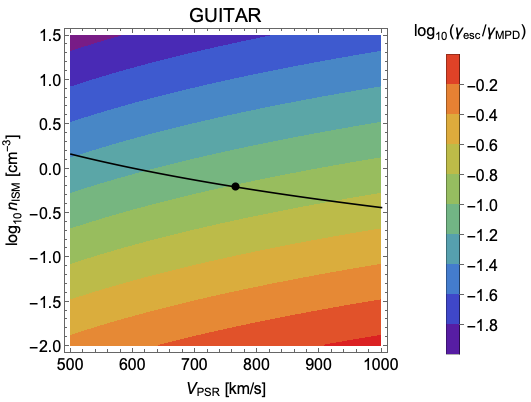}\,
\includegraphics[width=.445\textwidth]{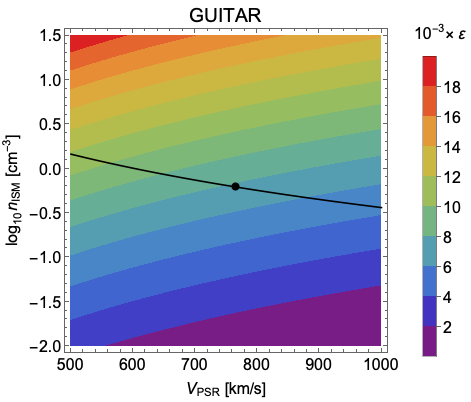}
\caption{
Guitar Nebula: Color map of the ratio, $\gmesc/\gmMPD$ (base-10 logarithm, left panel), and the efficiency, $\epsilon$ (right panel), as a function of the pulsar velocity, $\Vpsr$ (within the estimated uncertainty  \citep{deVries:2022}), and the ambient number density, \repo{$n_\mathrm{ISM}$}. The black curve shows the relation we used to estimate \repo{$n_\mathrm{ISM}$} (pressure equilibrium at the bow shock standoff distance, $d_0$). The black points indicate the position of the system for the best estimate of $\Vpsr$ (and \repo{$n_\mathrm{ISM}$}).}
\label{fig:guitFIGURE}	
\end{figure*}
The filament is $\sim 0.6\U{pc}$ long and shows a sharp leading edge (in the direction of motion of the pulsar) and a smoother trailing one, with hints of variation in the photon index \citep{Johnson:2010} suggestive of synchrotron cooling \citep{deVries:2022}.

When we apply our model to the Guitar Nebula feature, we find 
that the amplified magnetic field, $\Delta B\approx 80\U{\mu G}$, is produced by a current of particles, with $\gmesc\approx 0.1 \gamma_{\rm MPD}$, carrying a negligibly small fraction of the pulsar spin-down energy, $\eps\approx 6 \times 10^{-3}$. The amplified field also ensures that the Larmor radius of X-ray emitting particles, which we find to be a factor of $\sim 3$ more energetic than the current-carrying ones, is much smaller than the feature width, $R_L(\gmX)/w_f\approx 3 \times 10^{-3}$, which is consistent with our picture in which the latter is determined by synchrotron losses. 

A final consistency check concerns the relative size of the injection region and the Larmor radius of escaping particles in the unperturbed magnetic field. If $R_{L}(\gmesc,B_0)\gg R$, 
%the latter is much larger than $R$, 
the particles will occupy a much larger region of space than assumed in Eq.~\ref{eq:j} and the current density will decrease proportionally.
%be much smaller than we assumed. 
In fact, $R_L(\gmesc, B_0)=m\ c^2 \gmesc/(e B_0)\gg R$, but based on numerical simulations of the particle escape from these systems \citep{Olmi_Bucciantini:2019c}, the flux of particles leaving the nebula is well collimated, and limited to a surface area of $\sim d_0^2$.
%of order a fraction of $d_0^2$. 
Consistently, we find that the necessary value of $R$ is $R\approx d_0/2$ for the Guitar Nebula, implying a highly collimated initial particle flow, within an angle, $\alpha\sim R/R_L(\gmesc, B_0)\approx 3^\circ$. This initial anisotropy is quickly destroyed as the instability grows and full isotropization is also achieved for the more energetic X-ray emitting particles on a timescale,
$t_{\rm iso}\ll \tau_{\rm sync}$ (see App.~\ref{sec:app2}),
ensuring the validity of the assumptions underlying our calculations.
In Fig.~\ref{fig:guitFIGURE} we show how $\gmesc/\gmMPD$ and $\eps$ vary with \repo{$n_\mathrm{ISM}$ and $\Vpsr$, both depending on the uncertain estimate of the source distance.}
%the uncertain \repo{$n_\mathrm{ISM}$} and source distance.

\begin{figure*}
\centering
\includegraphics[width=.49\textwidth]{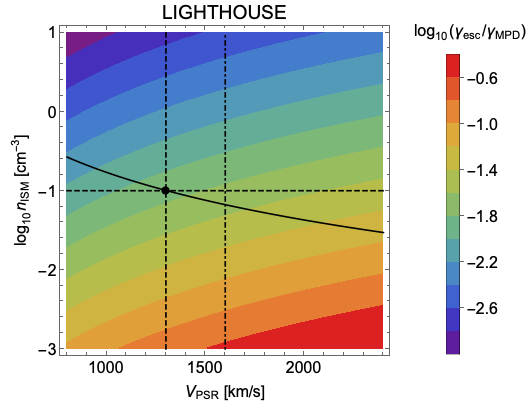}\,
\includegraphics[width=.455\textwidth]{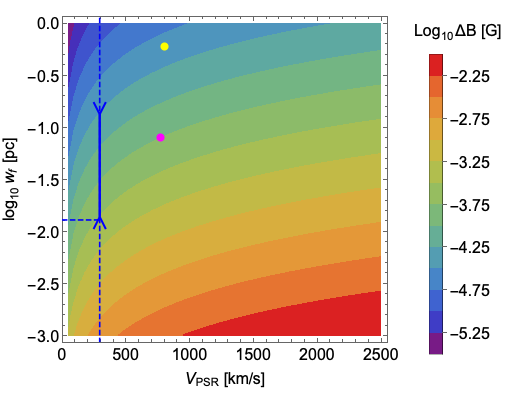}\,
\caption{%\LEt{***If possible, please insert a general title describing the figure as a whole in a telegraphic style that omits the initial articles (the, a, an), here and anywhere else where one is not provided for figure or table legends.}
{\it Left panel}: Color map of the ratio $\gmesc/\gmMPD$ (base-10 logarithm) for the Lighthouse Nebula, as a function of the pulsar velocity, $\Vpsr$, and ambient number density, \repo{$n_\mathrm{ISM}$}. The solid black curve shows the relation we used to estimate \repo{$n_\mathrm{ISM}$} (pressure equilibrium at the bow shock standoff distance, $d_0$). The black point indicates the position of the system based on the best estimate \citep{Pavan:2014} of \repo{$n_\mathrm{ISM}$}: \repo{$n_\mathrm{ISM}=0.1\;\mathrm{cm}^{-3}$}, $\Vpsr=1300\;\mathrm{km\,s}^{-1}$.
The vertical lines mark the maximum value of $\Vpsr$ that makes our model viable for \repo{$n_\mathrm{ISM}=0.1\;\mathrm{cm}^{-3}$} ($V_{\mathrm{psr,\,max}}=1300\;\mathrm{km\,s}^{-1}$, dashed line) and \repo{$n_\mathrm{ISM}$} determined from $\Vpsr$ and $d_0$ ($V_{\mathrm{psr,\,max}}=1600 \;\mathrm{km\,s}^{-1}$, dash-dotted line). \\
\textit{Right panel}: Color map of $\Dl B$ as a function of $\Vpsr$ and $w_{f}$ (uncertainties on the source distance reflect on both quantities). The best values for the Guitar and the Lighthouse Nebula are represented as magenta and yellow circles, respectively. The upper and lower limits obtained for J2030+4415 are shown as blue downward and upward arrows, respectively. The horizontal blue line shows the current filament thickness.
} 
\label{fig:BvsV}	
\end{figure*}

%%%%% LIGHTHOUSE %%%%%%%%%%%%%%%%%%
In light of the encouraging results obtained for the Guitar, we applied our model to the other two systems with well-surveyed X-ray filaments, the Lighthouse Nebula \citep{Pavan:2016,Klingler:2023} and PSR J2030+4415 \citep{deVries-J2030:2022}. The filament associated with the powerful Lighthouse Nebula is the only one also observed in hard X-rays \citep{Klingler:2023}, thanks to a recent NuSTAR campaign.
%(the measured flux is reported in Table~\ref{tab:infosystems}). 
The feature remains clearly visible up to $\sim 25$ keV, with a width and length compatible with those inferred from the higher-resolution Chandra data.
The measurement of $w_f$ along the feature is complicated by the observed striped morphology. Moreover, there is a large uncertainty on the pulsar speed, inferred from the association with a young ($10$--$30\U{kyr}$) supernova remnant in the vicinity \citep{garcia:2012}. 
The Lighthouse feature is the longest one currently known and the only one showing a bent morphology. In our model, this bending is to be understood as a result of the GMF structure, being the length of the feature not much smaller than the presumed GMF correlation length of $\sim$ tens pc.
%\LEt{***Please use “length by $\sim$ tens pc” if this is the difference between the two lengths and “length of $\sim$ tens pc” if this is the presumed GMF correlation length.}
%
Focusing on the Chandra band, we find our model to be consistent with observations of the Lighthouse only if $\Vpsr\leq 1300\; \mathrm{km\,s}^{-1}$, corresponding to $\Delta B\approx 26\, \mu\mathrm{G}$ and $\gamma_{\rm esc}\approx 8 \times 10^7 (\approx \gamma_{\,\mathrm{X}_1})$. A higher pulsar proper motion implies a higher magnetic field ($\Delta B\propto \Vpsr^{2/3}$ from Eq.\ref{eq:closing1}-\ref{eq:gesc}), and hence a lower $\gamma_{\,\mathrm{X}}$ ($\gamma_{\,\mathrm{X}} \propto \Vpsr^{-1/3}$). At the same time, our method for estimating the density based on the standoff distance implies \repo{a number density of the ISM}, $n_\mathrm{ISM}\propto \Vpsr^{-2}$, and $\gmesc\propto \Vpsr^{7/3}$ (Eq.~\ref{eq:gesc}), so that, as $\Vpsr$ increases, our assumption that $\gamma_{\,\mathrm{X}}\geq\gmesc$ is soon violated. If, rather than estimating \repo{$n_\mathrm{ISM}$} as described above, we assume that \citep{Pavan:2014} \repo{$n_\mathrm{ISM}=0.1 \; \mathrm{cm}^{-3}$}, then $\gmesc\propto \Vpsr^{4/3}$  and our model will work for\footnote{Notice however that values of $\Vpsr>1600\; \mathrm{km\,s}^{-1}$ are beyond $5\sigma$ of the peak velocity of the pulsar distribution \citep{FGK:2006} and are only suggested based on the possible association of PSR J1101–6101 with SNR G290.0–0.8} $\Vpsr\leq 1600 \; \mathrm{km\,s}^{-1}$. The range of validity of our model and the resulting value of $\gmesc/\gmMPD$ is shown in
% in the left panel of
Fig.~\ref{fig:BvsV}.
% We proceed as for the Guitar and limit our calculations to the {\it Chandra} band, where the structure is better resolved, and consider an intermediate pulsar speed of 1600 km/s. We find $\gmesc \simeq \times10^8\, \left( n_0/0.07\,\mathrm{cm}^{-3}\right)^{1/2} \left( E\rs{X,keV}/1.5\,\mathrm{keV}\right)^{-2/3}$ and $\Dl B \simeq 26\U{\mu G} \, \left( n_0/0.07\,\mathrm{cm}^{-3}\right)^{1/2} \left( E\rs{X,keV}/1.5\,\mathrm{keV}\right)^{-1/3}$, with an efficiency $\eps \simeq 2\times 10^{-2}$,  $\epslim\simeq 3\times 10^{-4}$ and $R\simeq 2 d_0$.
%
\begin{table*}%[ht]
\centering
\fontsize{10}{10}\selectfont
\begin{tabular}{|l|c|c|c|c|c|c|c|}
\hline
System & \repo{$n_\mathrm{ISM}$} &  $\gmMPD$ & $\gmesc$  & $\Dl B$ & $\eps$ & $R/d_0$ \\
 & cm$^{-3}$   &    &   &  $\mu$G  & $\times 10^{-3}$    &    \\
\hline
Guitar&  0.6 & $1.2\times 10^8$ & $1.2\times 10^7$ & 78 & 6  &  0.5 \\
\hline
Lighthouse & 0.07 & $4.0\times 10^9$  & $8\times 10^7$ & 26 & 19  &  2  \\
\hline
J2030+4415&  4.0 & $5\times 10^8$ & $\lesssim 4 \times 10^7 \left(L_f/\mathrm{pc}\right)$ & $40 < \Dl B <181$ & --  & --   \\
\hline
\end{tabular}
\caption{\label{tab:results}Summary of the results for the three filaments considered. Here, \repo{$n_\mathrm{ISM}$} is the ambient number density, $\gmMPD$ the Lorentz factor associated with the pulsar MPD, $\gmesc$ the Lorentz factor of the particles producing the instability, $\Dl B$ the amplified magnetic field in the filament, $\eps$ the fraction of the pulsar power carried by the escaping particles, and $R/d_0$ the radius of the filament base, in terms of the bow shock standoff distance, $d_0$. 
For J2030+4415 it is only possible to define a range for the amplified field, since there are indications that the filament formed only $t_\mathrm{event}\sim 32$ yrs ago and its present $w_f$ might not be determined by synchrotron losses. A lower limit comes from asserting that X-ray emission must be produced by particles with $\gm\leq \gmMPD$, while an upper limit is derived
%\LEt{***“while there is an upper limit”? The predicate is missing here. Please add what the upper limit is doing.} 
from the condition $t_\mathrm{event} < \tsync$ (see main text). From the maximum magnetic field, through Eq.~\ref{eq:closing1} one also determines an upper limit of $\gmesc$, which depends on $L_f$, since the feature might not be in its final form.}
\end{table*}

%%%%% J2030 %%%%%%%%%%%%%%%%%%
The very thin feature in J2030+4415 is powered by a $\gamma$-ray pulsar, with a spin-down luminosity corresponding to $\gmMPD=5\times 10^8$. Multi-epoch $\mathrm{H}_\alpha$ observations of the bow shock \citep{deVries-J2030:2022} clearly indicate an important rearrangement of the apex around $t_\mathrm{event}\simeq32 \, \mathrm{yr}$ ago, possibly due to a sudden variation in the ambient medium density, causing a compression of $d_0$. 
If the observed $w_f$ were determined by the synchrotron lifetime of the particles then, based on our model, we would estimate %$\Delta B\approx 160 \mu$G 
$\Delta B\approx 120\, \mu$G
and $\gmesc\approx 10^8$, which would imply that $\tsync\sim 60\,\, \mathrm{yr}> t_\mathrm{event}$. However, it is difficult to believe that the feature was formed before $t_\mathrm{event}$, so in this case we think that we are seeing all the particles that have been injected during $t_\mathrm{event}$ and $w_f=V_{\rm psr}\sin\theta_f \,t_{\rm event}$.
In this scenario, we can only estimate boundaries for $\Delta B$: the condition $t_{\rm event}<t_{\rm sync}$ implies that $\Delta B<181 \;\mu\mathrm{G}$, while the condition that X-ray emission must come from particles with $\gamma_{\,X}<\gmMPD$ implies that $\Delta B>40\,\mu\mathrm{G}$. These boundaries, reported in Table~\ref{tab:results}, clearly show that the field must also be amplified by at least a factor of $\sim 10$ in this case. The right panel of Fig.~\ref{fig:BvsV} shows the relation between the amplified magnetic field, the pulsar speed, and the filament thickness of all three sources.

%%%%%%%%%%%%%%%%%%%%%%%%%%%%%%%%%%%%%%%%%%
\section{Discussion and conclusions}
We claim that the filaments of non-thermal X-ray emission emerging from selected BSPWNe may be the first clear indication of the excitation of NRI due to a pencil beam current of charge-separated electrons (or positrons) leaving the parent nebulae. In two out of three cases, the length, thickness, and X-ray luminosity of the filaments can be well accounted for. In the third case, due to the complex history of the source, it is only possible to derive boundaries on the magnetic field, which however imply efficient amplification, possibly explained by particles leaving the source at an energy close to the pulsar potential drop.

The amplified field should not however hinder the ballistic motion of the leptons, at least in the beginning, so as to allow the particles to populate the whole length of the filament. This is exactly what is expected to happen when the beam excites the NRI \citep{Bell:2004}. In order for the mechanism to work, particles' injection into the ISM needs to be collimated within a narrow range of pitch angles. The filaments are predicted to follow the structure of the large scale GMF at the location of the BSPWN. 

The interpretation of the filaments in terms of excitation of the NRI by charge-separated electrons or positrons released by BSPWNe into the ISM is rich in implications. The amplified turbulence level can lead to extended confinement of the particles in selected regions around the sources, with implications for the pulsar's contribution to cosmic rays (CR)
%\LEt{***Acronyms should be spelled out upon first appearance in the abstract and then again beginning with the introduction. All subsequent appearances should be in the acronym form (unless at the beginning of a sentence).} 
leptons \citep{Schroer2023}.
In addition, it is tempting to speculate on a relation between the processes discussed here and the recently discovered phenomenon of TeV pulsar halos, regions of very high-energy ($\gtrsim \mathrm{TeV}$) gamma-ray emission extending for tens of pc around a few pulsars \citep{TeVHalos2017}. 
In the halos, too, the particle transport seems to be suppressed by two to three orders of magnitude \citep{TeVHaloRecv} compared with the Galactic values, as has been inferred from CR secondary and primary ratios \citep{Schroer2023,Evoli2020}, and the most obvious source of turbulence to explain the suppression seems to be the escaping particles themselves.

\repo{
The recently discovered radio filaments around some bow shock nebulae (see e.g., \citealt{Khabibullin:2024}) or the radio filaments observed in the Galactic center region \citep{Yusef-Zadeh:1984,Morris:1985,Yusef-Zadeh:1987,Meerkat:2023} may appear to be reminiscent of a similar process.
However, the observed radio emission requires particles of much lower energy than the X-ray one ($\gamma\sim 10^4-10^5$), and no need to amplify the ambient magnetic field. Moreover, those particles can more efficiently escape the bow shock from the tail rather than the head (due to their smaller Larmor radii, \citealt{Olmi_Bucciantini:2019c}), and then illuminate the preexisting structures of the magnetic field \citep{Barkov_Lyutikov:2019}.
%However the observed radio emission requires particles of much lower energy than the X-ray one ($\gamma\sim 10^4-10^5$), and no need to amplify the ambient magnetic field. Moreover those particles can more efficiently escape the bow shock from the tail rather than the head (due to their smaller Larmor radii, \citealt{Olmi_Bucciantini:2019c}), and then illuminate the pre-existing structures of the magnetic field \citep{Barkov_Lyutikov:2019}.
%
Hence, we argue that these phenomena are most likely of a different origin than the filaments around BSPWNe.
}

%%%%%%%%%%%%%%%%%%%%%%%%%%%%%%%%%%%%%%%%%%

\begin{acknowledgements}
This work has been partially funded by the European Union - Next Generation EU, through PRIN-MUR 2022TJW4EJ.
B. Olmi, E. Amato and R. Bandiera also acknowledge support from the Italian National Institute for Astrophysics with PRIN-INAF 2019.
\end{acknowledgements}

\bibliographystyle{aa}
\bibliography{biblio}

%TC:ignore
\begin{appendix}
\label{sec:methods}
%%%%%%%%%%%
\section{Initial collimation of the particle beam}\label{sec:app1}
In the main text, we have shown that a good collimation -- that is, small pitch angles -- is one of the requirements for the formation of the observed filaments.
A good collimation is also a natural outcome of the fact that the particles move from a region with a high magnetic field (the bow shock head),
%%%
\begin{eqnarray}
    B\rs{head} &\simeq &\sqrt{24\pi(1-\eta)\rho_\mathrm{ISM}}\;\Vpsr \simeq \\ \nonumber
    &\simeq&
    400\sqrt{1-\eta}\sqrt{\frac{n_\mathrm{ISM}}{0.5\U{cm^{-3}}}}\left(\frac{\Vpsr}{500\U{km\,s^{-1}}}\right)\U{\mu G}\,,
\end{eqnarray}
%%%
to a typical ISM magnetic field, around $3\U{\mu G}$ in magnitude. 
It is well known that, for particles with a small Larmor radius compared to the scale length of $B$ variation, the enclosed magnetic flux is an adiabatic invariant and leads to the constancy of $B\sin^2\al$.
Let us assume that, originally, the pitch angles of the escaping particles uniformly cover a $2\pi$ solid angle, equivalent to a mean value of $\sin\al=\pi/4$. This would imply, for the reference values given above, an average $\sin\al\simeq0.07$.
For the considered cases, the ``small Larmor radius'' approximation is valid, being the ratio of the gyro radius of the particles with energy $\gmesc$ in the magnetic field of the bow shock head, $0.1-0.3\,d_0\,(B\rs{head}/100\U{\mu G})^{-1}$, where the bow shock standoff distance, $d_0$, gives the scale of the variation in the field from the inside to the outside.
%%%%%%%%%%
\section{Delayed isotropization of X-ray emitting particles in the feature}\label{sec:app2}
Once the instability is saturated, the particles with energy, $\gmesc$, start to diffuse in the amplified magnetic field. 
One might wonder how much later this will also happen to the particles responsible for the emission at the X-rays, due to the fact that they have a larger Lorentz factor.

During its motion in the perturbed magnetic field, $\Dl B$, a particle with Lorentz factor $\gamma_{\,\mathrm{X}}>\gmesc$ will be deflected by an angle, $\Delta\theta_i=l/R_L^*(\gamma_{\,\mathrm{X}})$, on the interaction scale $l=1/\kmax^*$ and where $R_L^*(\gamma_{\,\mathrm{X}})$ is the Larmor radius of the particle in the amplified magnetic field. In a distance, $L$, the number of interactions (or deviations) that the particles experience is $L/l$.
Then the total deflection can be estimated as $\Delta\theta=(L/l)^{1/2}\, l/R_L^*(\gamma_{\,\mathrm{X}})$.
The isotropization length ($L\rs{iso}$), or the isotropization timescale ($t\rs{iso}\simeq L\rs{iso}/c$) is obtained by requiring that $\Delta\theta=1$, so that
\begin{eqnarray}\label{eq:isoscale}
    t\rs{iso} &\simeq &\frac{L\rs{iso}}{c} = \frac{1}{\Omega_c}\frac{\gmX^{\,2}}{\gmesc}\left(\frac{\Dl B}{B_0}\right)^{-1} \simeq \\ \nonumber
    &\simeq &  0.01\,\U{yr}\; \left( \frac{\gmesc}{10^7} \right)^{-1}  \left( \frac{\Delta B}{50 \U{\mu G}}\right)^{-2}  \left( \frac{E\rs{X,keV}}{0.5\U{keV}}\right)\,.
\end{eqnarray}

With the typical lifetime of X-ray emitting particles in the amplified field being much longer than $t\rs{iso}$ -- namely, $\tau\rs{synch} \simeq 220\U{yr} \;(\Delta B/50\U{\mu G})^{-3/2} \, (E\rs{ph,keV}/0.5\U{keV})^{-1/2}$ -- the distribution of particles can be safely assumed to be isotropic for the sake of  computing the observed X-ray emission.
\end{appendix}

%
%TC:endignore
\end{document}